# Buffer-less Gallium Nitride High Electron

# Mobility Heterostructures on Silicon

*Saptarsi Ghosh[1,2,*], Martin Frentrup[1], Alexander M. Hinz[1], James W. Pomeroy[3], Daniel Field[3],*

*David J. Wallis[1,4], Martin Kuball[3], and Rachel A. Oliver[1]*

[1]Department of Materials Science and Metallurgy, University of Cambridge, Cambridge CB3 0FS, United Kingdom

[2]Department of Electronic and Electrical Engineering, Swansea University, Swansea SA1 8EN, United Kingdom

[3]H.H. Wills Physics Laboratory, University of Bristol, Bristol BS8 1TL, United Kingdom

[4]Centre for High Frequency Engineering, Cardiff University, Cardiff CF24 3AA, United Kingdom

*E-mail: saptarsi.ghosh@swansea.ac.uk

**Abstract:**

Thick metamorphic buffers are perceived to be indispensable for the heteroepitaxial integration of III-V semiconductors on silicon substrates with large thermal expansion and lattice mismatches. However, III-nitride buffers in conventional GaN-on-Si high electron mobility transistor (HEMT) heterostructures impose a substantial thermal resistance, throttling heat extraction, which reduces device efficiency and lifetime. Herein, bypassing the buffer, we demonstrate the direct growth of GaN after the AlN nucleation layer on silicon by metal-organic vapor phase epitaxy (MOVPE). By varying reactor pressure, we modulate the growth stress in the submicron epilayers and realise threading dislocation densities similar to that in thick buffered structures. We achieve a GaN-to-substrate thermal resistance of $(11 \pm 4)$ m$^2$ K GW$^{-1}$, an order of magnitude reduction over conventional designs on silicon and one of the lowest on any non-native substrate. AlGaN/AlN/GaN heterojunctions on this platform show a characteristic 2D electron gas (2DEG), the room-temperature Hall-effect mobility of which, at over 2000 cm$^2$/V-s, rivals the best-reported values. The low-temperature magnetoresistance of this 2DEG shows clear Shubnikov-de-Haas oscillations, a quantum lifetime $> 0.180$ ps, and tell-tale signatures of spin-splitting. These results may establish a new paradigm for nitride HEMTs, potentially accelerating applications from energy-efficient transistors to fundamental investigations on electron dynamics in this 2D wide-bandgap system.

**Introduction:**

The success of semiconductor devices behind the quantum revolution is intricately related to long term developments in epitaxial techniques including the control of material defects. For example, buffer optimisation[1] and acceptor activation[2] were pivotal for the growth of InGaN/GaN quantum well (QW) based light-emitting diodes (LEDs) by metal-organic vapour phase epitaxy (MOVPE) and ushered in the era of energy-efficient lighting[3]. From the same III-nitride family, heterostructures of GaN and its alloys with AlN are now a frontrunner for high-frequency, high-power microelectronics by harnessing these materials' wide bandgap (3.4 eV to 6.0 eV), large breakdown field (> 3 MV/cm), high electron saturation velocity ($\geq 3 \times 10^7$ cm/s) along with robust thermal and chemical stability. At the AlGaN/GaN interface in metal-polar heterostructures, the combination of a lack of inversion symmetry along [0001] and a strong polarity of the metal-nitrogen bonds leads to a polarization-induced highly-mobile 2D electron gas[4] (2DEG). Its carrier density is an order of magnitude higher than modulation-doped AlGaAs/GaAs 2DEGs, and its conductivity is superior to electron channels in competing wide bandgap SiC, $Ga_2O_3$, and diamond materials[5]. High electron mobility transistors (HEMTs) based on such dopant-free 2DEGs in III-nitrides underpin 5G radio-frequency amplifiers[6], power converters[7], extreme environment electronics[8], chemical sensors[9], THz detectors[10], thermoelectric harvesters[11], and flexible electronics[12] applications.

However, with one of the highest heat fluxes[13] for solid-state transistors, a major roadblock for high power density GaN HEMT circuits is the extraction of heat. High operating temperatures from localised Joule heating in the gate-drain region not only degrade transconductance but also exponentially increase the probability of device failure[14]. Hence, all thermal resistances between the channel and heat-sink need minimisation. This is particularly disadvantageous in terms of the application of thermally-resistive sapphire (thermal conductivity, $k \approx 40$ W/m·K), the typical manufacturing platform for nitride LEDs, as a substrate for heteroepitaxial HEMTs. Instead, hexagonal SiC ($k \approx 400$ W/m·K) is considered the optimal substrate for GaN RF HEMTs as economical bulk nitride wafers are still unavailable. AlN is the commonly used nucleation layer (NL) for nitride epitaxy on non-native substrates such as sapphire[15], SiC[16], and Si[17]. SiC possess only ≈1% basal-plane lattice parameter mismatch with AlN. Hence, for SiC, low dislocation-density GaN on an AlN NL can be achieved without needing additional defect-filtering layers, and this design keeps the channel-to-substrate aggregate thermal resistance low[18]. Nonetheless, high costs, the existence of export restrictions, and poor availability of large-diameter SiC have long-prompted the urge for their replacement with inexpensive large-area silicon ($k \approx 150$ W/m·K) substrates, that can also leverage widely-available technological know-how on Si-based processing and legacy CMOS fabrication facilities.

Lithography-free epitaxial integration of III-V semiconductors on silicon wafers has traditionally relied upon buffers between the functional layers and the substrate[19,20]. Specifically, GaN grown on Si experiences significant tensile stress during post-growth cooldown owing to the ≈ 50% mismatch in coefficients of thermal expansion (CTE) between substrate and epilayer, often cracking the epilayers. In addition, the ≈19% lattice mismatch between AlN and Si generates significantly more threading dislocations (TDs) in the NL, which in-turn propagate into the GaN. To tackle these issues, thick compositionally graded-AlGaN[17,21] or AlN/GaN superlattice layers[22] on top of the NL have been integral components of GaN-on-Si HEMTs for the last two decades. These buffers induce epitaxial compressive stress in GaN to balance the post-growth tensile stress and partially annihilate TDs generated in the NLs[21,22]. On the flip side, such nitride buffers prolong the overall growth duration, increasing the thermal budget and the usage of unsustainably-produced ammonia and hydrogen carrier gas. This is contrary to the objectives of greener manufacturing as energy, financial, and environmental costs of semiconductor production are becoming a major concern[23]. More importantly, heat-carrying phonons suffer significant alloy scattering in $Al_xGa_{1-x}N$ (for $0.1 < x < 0.9$) and interface scattering in short-period AlN/GaN superlattices. Both of these buffers possess very-low thermal conductivity ($k \approx 10$ W/m·K) [24,25] compared to either AlN[26] ($k \approx 340$ W/m·K) or GaN[27] ($k \approx 140$ W/m·K) and appear as large thermal resistance in series. This proportionately raises the device temperature[28], causing operational and reliability concerns. In fact, even after substrate thinning, self-heating still lowers the RF output of HEMTs on Si at higher powers[29]. Compared to benchmark results on SiC[30], today's commercial GaN-on-Si devices are intentionally derated by an order of magnitude. Thus, bypassing the buffer bottleneck without sacrificing the stress-management and structural benefits they enable has become an immediate need to unleash the potential of nitride HEMTs.

With or without a buffer, offsetting the post-growth tensile stress ($\approx 1$ GPa for $\approx 1000°C$ growth temperature[31]) necessitates inducing compressive stress of similar magnitude during the GaN growth. Notably, AlN's smaller in-plane lattice constant (3.112 Å) compared to GaN (3.189 Å) itself entails about 11 GPa of compressive stress. Hence, at least theoretically, countering the tensile stress is plausible even for direct growth on a relaxed AlN nucleation layer (NL) if just $\approx 10\%$ of this compressive-stress source can be retained during growth. Yet, numerous, mostly early, attempts[32,33,34,35] to grow buffer-less large-area GaN-on-Si have culminated in cracks after cooldown, suggesting a failure to control the compressive stress relaxation. Available *in situ* analysis[32,36] showed that without buffers, the stress in GaN was either tensile from the beginning, or if compressive, decayed within the first $\approx 100$ nm, and subsequent growth occurred in the tensile regime. In the last decade, only sporadic reports[37,38,39,40,41,42,43,44] of crack-free buffer-less GaN-on-Si have emerged and these neither provide quantifiable insight into their stress-structure evolution nor evidence of simultaneous optimisation of electronic and thermal properties. Thus, the quantitative understanding needed for reproducible synthesis and optimisation of buffer-less epi-structures along with the immediate performance advantage for their realisation, is unavailable.

**Results and Discussion:**

**Modulation of stress relaxation with growth parameter:**

In this work, we designed the individual epilayer thicknesses for buffer-less GaN/AlN/Si heterostructures based on strain and thermal considerations. It is known that for the AlN nucleation layer, increasing its thickness ($t_{AlN}$) reduces its dislocation density[45]. However, since the strain state of an AlN NL on Si is constantly tensile during growth, this also increases the strain energy of this layer and, above 200 nm, the potential for cracking during growth[36]. Hence, as a trade-off, $t_{AlN}$ was kept at $\approx 150$ nm in our experiments. In terms of the GaN thickness ($t_{GaN}$), simulations can predict the optimum thickness for thermal performance. For GaN HEMTs with a silicon substrate, the aggregate thermal resistance initially decreases with increasing GaN thickness ($t_{GaN}$) and then plateaus after $\approx 500$ nm[46]. For HEMT membranes (i.e. structures where all layers underneath the GaN are etched away) integrated with synthetic diamond heat-spreaders, simulations[46] suggest little dividend for increasing $t_{GaN}$ above 1 μm. Large GaN thicknesses are not beneficial from the stress perspective either. Even on graded

AlGaN or superlattice buffers the initial compressive stress in the GaN gradually relaxes as the layer thickness increases[36,17], and hence with increasing thickness the cumulative compressive stress will eventually become lower in magnitude than the post-growth tensile stress. Accounting for all of these factors, we designed $t_{GaN}$ to be ≈ 750 nm to 800 nm, with the total epi-thickness of ≤ 1 μm.

For all growths, *in situ* removal of the native oxide on the silicon substrates preceded the growth of the AlN NLs, and both these steps were identical for all samples (see Supplementary Fig. 1). On these AlN NLs, GaN was directly grown after the change of growth conditions (see methods). For different samples the reactor pressure ($P_g$) during GaN growth was varied over an order of magnitude (200, 125, 75, 37.5, and 18 Torr), with all other parameters kept constant. The cross-sectional secondary electron micrograph (SEM) of an as-grown structure (Fig. 1a) shows the thin GaN epilayer on top of the NL.

During heteroepitaxy, the wafer-curvature changes due to the stress in the growing layer. Previously, for GaN/step-graded AlGaN grown on silicon, this real-time change in wafer-curvature (which is directly proportional to the layer's stress-thickness, $\sigma \cdot t$) can provide valuable insight into stress evolution[17]. Using the same formalism, the calculated stress-thickness versus thickness data for these GaN layers grown at different pressures is shown in Fig. 1b. The differential slope of these curves is proportional to the instantaneous stress ($\sigma_i$) and the slope from the origin is proportional to the cumulative mean-stress ($\sigma_{mean}$) at that thickness for the growing GaN layer. The instantaneous stress should be compressive (i.e. $\sigma \cdot t$ versus thickness should be negatively sloped) from the beginning for coherent growth on AlN. The data indicate that at higher pressures, the compressive growth regime does not start from the beginning, but the thickness for its onset is progressively shortened as $P_g$ is reduced. This is further confirmed from the evolution of $\sigma_{mean}$ with thickness (inset to Fig. 1b). For 18 Torr, the initial $\sigma_{mean}$ has the highest magnitude, and it continuously reduces during growth. This indicates that the successive atomic layers grow with increasingly relaxed lattice constant, in effect continuously decreasing the compressive stress averaged across the entire thickness[17]. In contrast, for 200 Torr, the layer becomes marginally more compressive in the initial phase (albeit with a much smaller value), followed by a regime of slow relaxation.

Intermediate behaviour is seen for pressures in between 18 Torr and 200 Torr, revealing that the decay of $\sigma_{mean}$ is never large enough to become tensile for any structure. Consequently, as shown in Fig. 1c, the final mean-stress of the GaN layers at the growth temperature systematically changes from (-0.41 ± 0.02) GPa to (-0.90 ± 0.01) GPa with the reduction in $P_g$. Except for the growth at 200 Torr, these mean-stresses were sufficient to prevent cracking (see Supplementary Fig. 2) by countering the subsequent tensile stress during cooldown.

Along with avoiding cracking, high-throughput automated fabrication requires the wafer-bow to be ≤ 50 μm. For a multilayer thin film, the wafer-bow at room-temperature (RT) is inversely proportional to the wafer-curvature (κ), which, in turn, depends on each layer's residual stress as $\kappa = \sum \Delta\kappa_i = \sum f\{(\sigma_{i\_growth} + \sigma_{i\_thermal}) \times thickness_i\}$. Accordingly, using the mean growth-stresses after 800 nm GaN growth ($\sigma_{i\_growth}$) from Fig. 1c and considering tensile thermal stress values ($\sigma_{i\_thermal}$) from the literature[31], the RT bows were calculated. As plotted in Fig. 1d, predicted values closely agree with the measured data, showing that simply controlling the GaN growth stress enabled the tuning of the post-growth wafer bow. This confirms that with a wafer bow of (48 ± 10) μm for $P_g$ = 75 Torr, and lower bows for lower $P_g$ values, the majority of the buffer-less GaN-on-Si structures of this series comply with the requirements for batch processing in CMOS fabs. In addition, shedding the buffer layers from the growth sequence provided substantial reduction in energy, material, and runtime (see Supplementary Fig. 3).

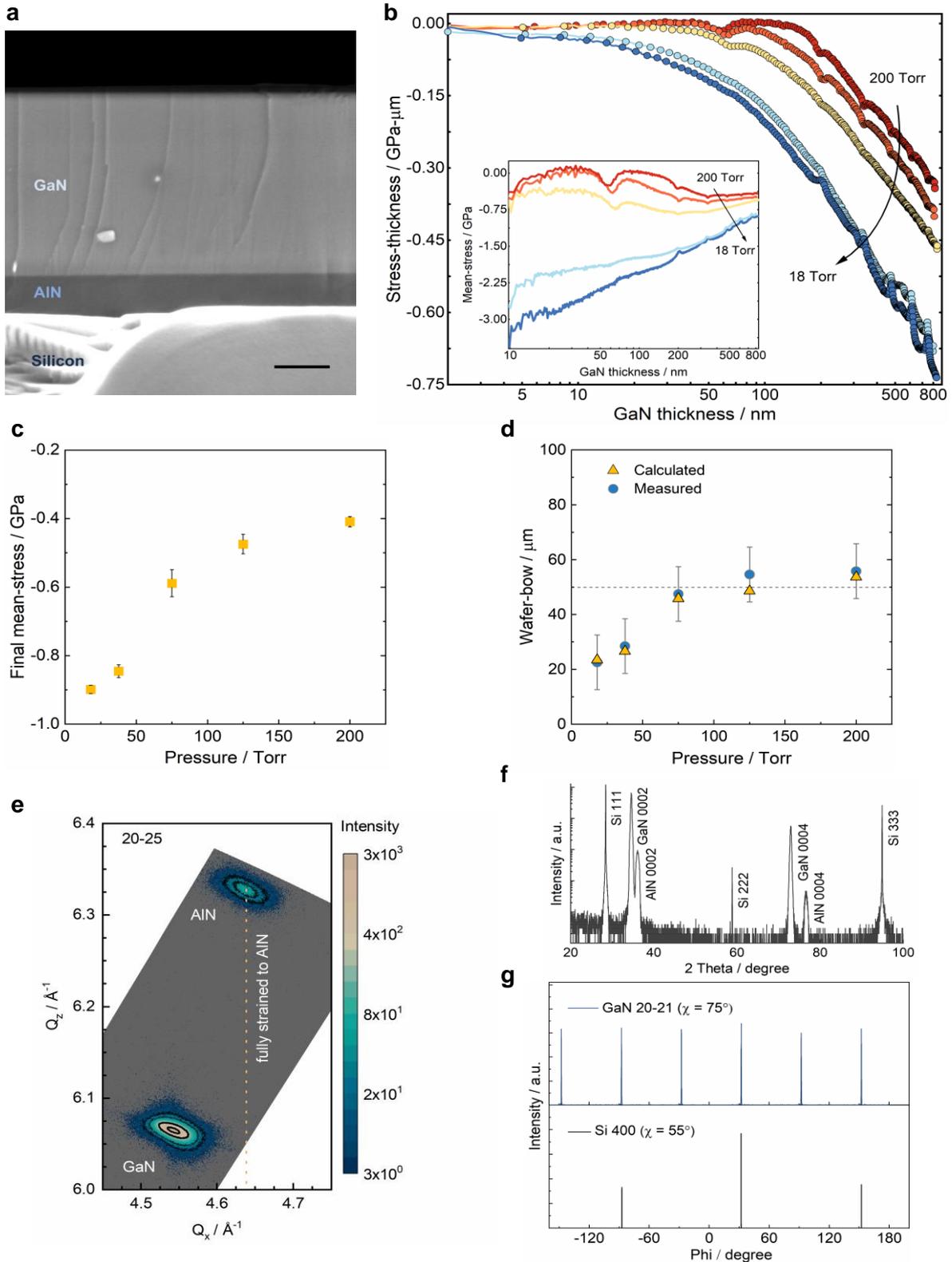

Figure 1. Stress and structural characterisation of GaN/AlN/Si heterostructures. (a) Cross-sectional SEM of a representative structure, scale bar is 200 nm. (b) Evolution of stress-thickness with thickness during the GaN growth at 200, 125, 75, 37.5, and 18 Torr. Inset shows the corresponding mean-stresses with increasing thickness. Note that the absolute stress-thicknesses after the AlN growth (i.e. prior to the GaN growth) were similar for all the growths, and in this image, these values have been offset to zero for comparison. (c) Mean-stresses accumulated after 800 nm GaN growth, averaged from multiple runs. (d) Post-growth measured (with a ± 10 μm resolution of the set-up) and calculated wafer bows at RT. (e) XRDRSM around the AlN 20-25 reflection. (f) On-axis ω-2θ scan showing peaks from the epilayer and substrate. (g) 360° φ-scans for peaks from GaN(20-21) and Si(400) planes.

Fig. 1e shows an X-ray diffraction (XRD) reciprocal-space map (RSM) around the asymmetric 20-25 AlN reflection for the epi-structure with the GaN layer grown at 75 Torr. Only reflections corresponding to AlN and GaN are observed, with no evidence observable for any ternary phase. Here, a smaller centroid $Q_x$ of the GaN peak compared to AlN indicates a larger average in-plane lattice constant, consistent with the gradual relaxation observed during growth. For the same structure, the XRD ω-2θ scan in Fig. 1f features only 000l peaks of the nitrides and 111 and higher order peaks of the silicon substrate, establishing GaN[0001] II Si[111]. Furthermore, only six sharp peaks are seen in the 360° azimuthal scan of the skew-symmetric reflection of GaN (Fig. 1g), confirming a single domain rotational alignment of the hexagonal unit cells on the cubic substrate. Altogether, these scans prove that along with stress-balancing, epitaxially preserving a sixfold atom arrangement of wurtzite nitrides on the threefold symmetric Si(111) surface with the wanted orientation of GaN[11-20]IISi[-110] does not require any buffers.

We subsequently considered the growth regimes to identify the mechanism behind the difference in strain relaxation among different structures. Figure 2a shows the reflectance data acquired during the growth of the GaN layers for the highest and lowest pressures. Fabry-Perot oscillations are visible in both, arising from the interference of the beams reflected from the GaN surface and the AlN/Si interface beneath. For $P_g$ = 18 Torr, the peak-to-peak magnitudes are the same for all periods. However, for the highest pressure, the initial sharp reduction is followed by gradual recovery (see Supplementary Fig. 4 for the intermediate behaviours at $P_g$ = 37.5 Torr to 125 Torr). To examine the corresponding evolution of topography at different stages, additional growth runs were terminated at the thicknesses marked in Fig. 2a. As seen in the AFM image in Fig. 2b, by point I after nominally ≈50 nm of growth at 18 torrs, the GaN layer is already continuous and completely covers the NL. This flat morphology persists as growth progresses (point II, ≈150 nm nominal thickness), along with a marginal increase in the lateral dimensions of the features (Fig. 2c). In contrast, isolated islands dominate the morphology at point I' for the growth at 200 Torr (Fig. 2d). At this pressure, even at point II', a considerable fraction of the layer is still uncoalesced (represented by dark areas in Fig. 2e). The merger of the islands requires a further progression of growth, as seen in Fig. 2f. This explains the reflectance evolution seen in Fig. 2a as the side facets of the islands would have diffusely scattered the incident laser beam causing a diminished detected intensity in the initial phase. Nonetheless, all the surfaces were smooth at the final layer thickness, with the sub-nm roughness required for abrupt heterojunctions (see Supplementary Fig. 5).

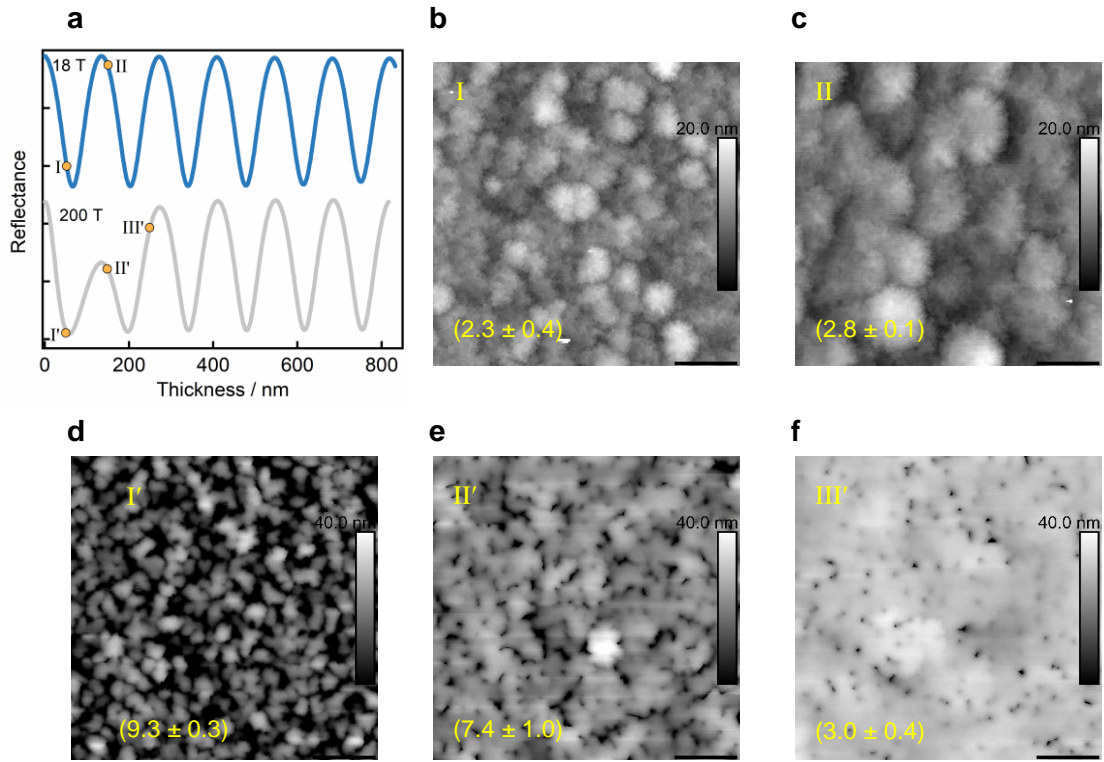

Figure 2. Real-time reflectance transients and their correlation with surface topography of the GaN layer. (a) Oscillations in reflectance transients acquired during 800 nm GaN growths at 18 and 200 torrs, respectively. Note that the data has been offset in intensity (without any multiplication) for clarity. Marked points denote the stage at which additional growth runs were terminated to inspect surface topography. (b) and (c) are 20 μm × 20 μm AFM images at different stage for growths at 18 torrs, whereas (d), (e), and (f) are AFM images for GaN grown at 200 torrs. The z-scale for (b)-(c) is 20 nm, and for (d)-(f) is 40 nm. The mean rms roughness (in nm) from three positions on the wafer is annotated for each AFM scan. The scale bar in (b)-(f) is 4 μm.

To estimate the threading dislocation density in these [0001] oriented epilayers, full-width at half maxima (FWHM) of HRXRD ω-scans for skew-symmetric 20-21 and on-axis 0002 reflections were assessed next. The measured FWHM values in Fig. 3a show that as $P_g$ increases, the 20-21 peak FWHM gradually reduces from 1073 arcsecs to 771 arcsecs. However, after an initial reduction from 648 arcsecs to 489 arcsecs, the 0002 peak FWHM does not decrease further for pressures higher than 75 Torr. Qualitatively, lower peak broadening indicates smaller in-plane twist and better out-of-plane tilt alignment among the individual crystallites, respectively. Quantitatively, from the 20-21 peak FWHM, the density of dislocations with edge-character (pure-edge and mixed), which are the most prevalent in wurtzite nitrides (usually < 2% are pure-screw types), can be estimated[47] as $D_e = \frac{(FWHM)^2}{4.35\, b_{edge}}$. This suggests an anticorrelated change in the edge-type dislocation density from 6×10⁹ cm⁻² to 3×10⁹ cm⁻² with the investigated growth pressures. Note that the XRD signal is collected from the entire GaN layer, and the calculated values represent a weighted average across the thickness. Among these, dislocations threading to the surface will directly affect the functionality of the subsequently grown HEMT channel layer. To quantify the TD density at the surface of these templates and distinguish different dislocation types, pits created by the surface terminations of dislocations were imaged by AFM[47] (see Supplementary Fig. 6). The scan results are summarised in Fig. 3b. A general trend of a reduction in the total dislocation density from $(4.9 \pm 0.7) \times 10^9$ cm$^{-2}$ to $(3.4 \pm 0.9) \times 10^9$ cm$^{-2}$ with increase in $P_g$ can be confirmed with pure edge dislocations accounting for 59% to 66% of the total densities. It must be emphasised that these dislocation densities are very similar to those achieved for GaN grown with different thick buffers[21,22,17].

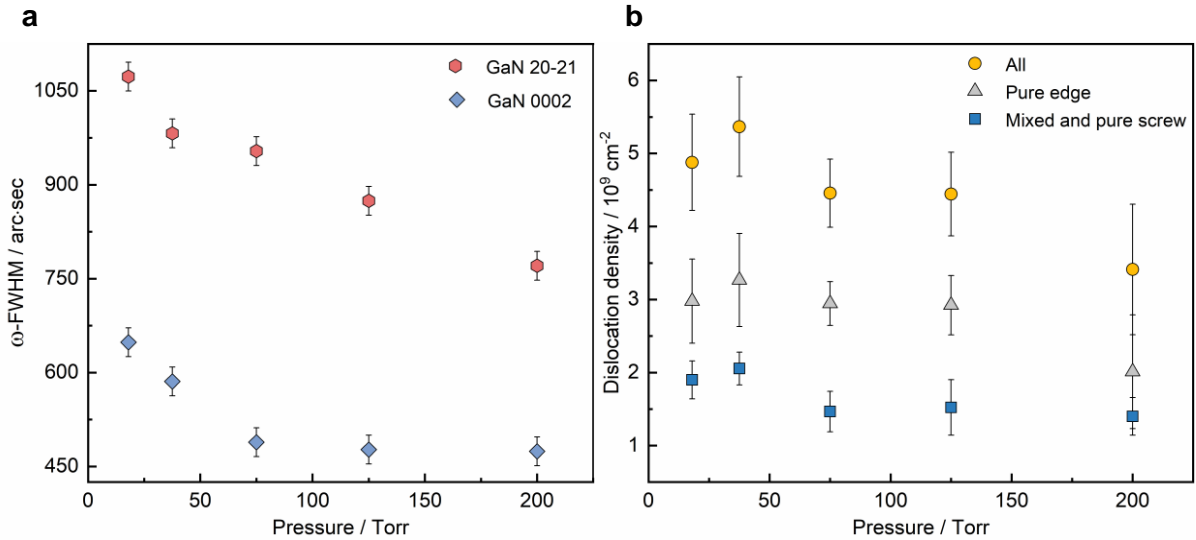

Figure 3. Dependence of structural properties of the GaN layer on growth pressure. (a) Variation in FWHM of HRXRD skew-symmetric 20-21 and symmetric 0002 ω-scan peaks of the GaN layer with growth pressure. The error bars represent the intrinsic broadening of the goniometer. (b) Variation in the density of the total and individual dislocation density at the surface of the GaN layers grown at different pressures. The error bars represent the standard deviations calculated from multiple positions in the same wafer.

These observations open up multiple avenues to control the involved stress evolutions. The 2.5% lattice mismatch between GaN and AlN provides a driving force for GaN to nucleate and grow as discrete islands. From Fig. 2a, it is evident that the duration of this island-mode growth increases with increasing reactor pressure. Also, little change in stress-thickness before island coalescence (in Fig. 1b) suggest that the tensile stress associated with coalescence itself (for the observed domain sizes, a maximum theoretical value on the order of GPa[48] can be estimated for GaN's modulus) largely compensates the compressive stress arising from the lattice mismatched growth during this period. Post-coalescence, the observed gradual decay of compressive stress is characteristically similar to that often associated with layer-by-layer growth of GaN on AlGaN and originates from dislocation climb driven by compressive stress.[36] While dislocation climb facilitates annihilation of dislocations with opposite Burgers vectors, for our structures, this mechanism is active only during the post-coalescence compressively-strained phase. Hence, it cannot be responsible for the reduction in dislocation density at higher growth pressures with increasingly delayed coalescence. Deliberate islanding has been key to achieving low dislocation density in GaN directly grown on sapphire. This acts by bending dislocations towards the facets which minimise the systems' free energy.[49] It is likely that a similar mechanism is involved in our samples up until coalescence, instead of climb. This appears to be the mechanism for reducing dislocations albeit at the expense of accumulating less compressive stress during GaN growth. Importantly, while the reactor pressure was implemented herein to modulate the coalescence, we note that recently both Lee et al.[43] and Zhan et al.[44] have successfully grown crack-free buffer-less GaN-on-Si with AlN NL by using a very large V/III ratio ($\geq$ 15000) during the initial growth phase. In fact, parameters like the V/III ratio, pressure, temperature, and growth rate are known control parameters used during the 3D-2D transformation of low-defect GaN-on-sapphire growth. In future, systematic exploration of all these correlated parameters during GaN on AlN NL growth can be expected to further expand the growth window for buffer-less nitrides on silicon.

**Analysis of thermal resistance for buffer-less GaN-on-Si:**

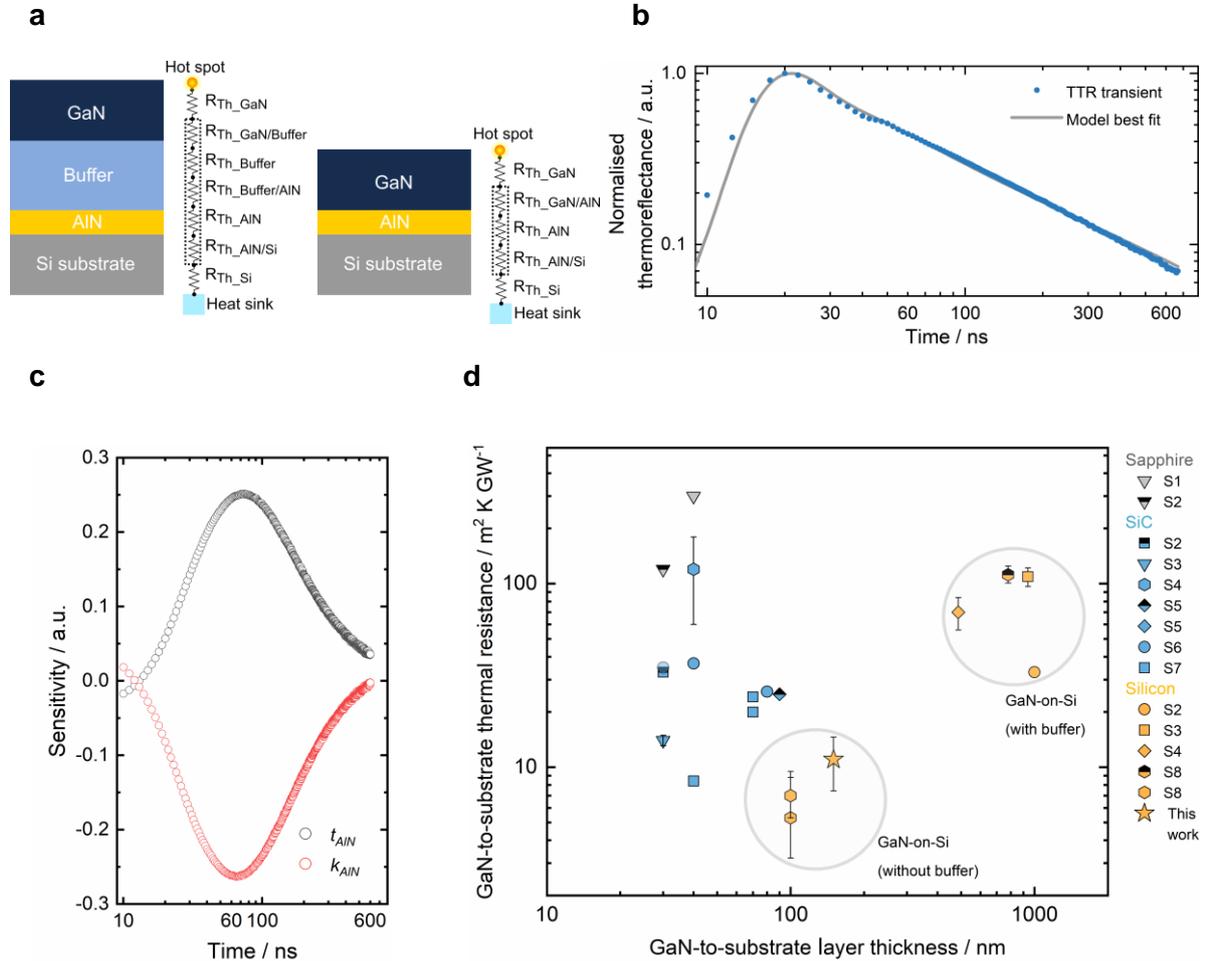

Figure 4. Thermal characterisation of GaN/AlN/Si heterostructures. (a) Schematic of the thermal resistances in the heat diffusion path from the channel hotspot to the heat-sink under the silicon. The individual resistances between the GaN and the Si substrate are lumped, which shows our bufferless design possesses fewer thermal elements than the conventional one. The interface resistances will likely be dominated by disorder rather than the phonon density mismatch in both. (b) Representative measured TTR trace for the sample for which the GaN layer was grown at 75 torrs, normalised to the peak thermoreflectance. Modelled best fit (solid line) closely approximates the experimental data. (c) Sensitivity analysis with respect to the variation in $k_{AlN}$ and $t_{AlN}$. (d) Comparing literature reports across heteroepitaxial GaN on sapphire, SiC, and Si, the measured sample has one of the lowest thermal resistances between the GaN layer and the non-native substrate (references for the data are listed in Supplementary Table 1).

For our sample series, we suggest that the GaN layer grown at 75 Torr presents the best balance of stress-management and micro-structural properties and experiments henceforth refer to GaN grown at this pressure. To evaluate the thermal transport across this GaN/AlN/Si heterostructure we used the nanosecond transient thermoreflectance (TTR) technique[50,27,51]. This non-invasive method involves a nanosecond pulsed laser to heat the sample surface and a continuous probe laser to measure the subsequent evolution of sample temperature with constant probing of surface Fresnel reflection (see methods). In multi-layered structures, this evolution depends on the thermal diffusivity of each layer through which the heat diffuses and possible thermal boundary resistances (TBR) at the interfaces between different materials. For our structure (Fig. 4a), the thermal resistances associated with the AlN layer are the only unknown parameters in the cumulative thermal resistance between GaN and substrate. We treat these as a lumped thermal boundary resistance ($TBR_{eff}$)[18], summing the AlN layer thermal resistance and the GaN/AlN and AlN/Si interface TBRs, $TBR_{eff} = \frac{t_{AlN}}{k_{AlN}} + TBR_{GaN/AlN} + TBR_{AlN/Si}$, where $t_{AlN}$ and $k_{AlN}$ are AlN layer thickness and thermal conductivity, respectively. For thin layers, it is impossible to accurately separate the intrinsic layer thermal resistance from the TBR contribution, so instead, we use an effective value for the AlN layer ($k_{AlN\_eff}$), which includes the TBR contribution, i.e. $TBR_{eff} = \frac{t_{AlN}}{k_{AlN\_eff}}$.

TTR transients were acquired from four different positions on the sample, and a representative normalised curve is shown in Fig. 4b. The experimental data were least-squared fitted with an analytical transmission-line axis-symmetric heat diffusion model[52] with $k_{AlN\_eff}$ as the fitted variable (see methods). Fig. 4b shows the simulated best fit for the representative transient. The sensitivity ($S_x$) of the thermoreflectance ($R$) to the fitting parameters ($x$) was also calculated and this is plotted in Fig. 4c where $S_x = \frac{\partial\{\ln(R)\}}{\partial\{\ln(x)\}}$. A high sensitivity to both $k_{AlN\_eff}$ and $t_{AlN}$ can be seen in the $\approx 20$ ns to 300 ns period. However, as the uncertainty in calibrated $t_{AlN}$ is minimal, this gives confidence in the extracted $k_{AlN\_eff}$. Based on the 150 nm thick AlN, we calculated the (mean ± std. dev.) $TBR_{eff}$ for this sample to be (11 ± 4) m$^2$ K GW$^{-1}$. Theoretically, the sum of the GaN/AlN[53] and AlN/Si[26] interfacial resistances (predicted by the diffuse mismatch model (DMM)) and the AlN layer thermal resistance (considering its bulk conductivity[26]) is predicted to be $\approx 2$ m$^2$ K GW$^{-1}$. However, in practice, the AlN's large dislocation density and limited thickness (compared to phonons with longer mean free path) will significantly lower its thermal conductivity compared to bulk material. Altogether, the best achievable $TBR_{eff}$ can be expected to be higher than the theoretical value. Nonetheless, the $TBR_{eff}$ in our sample is low enough to ensure a negligible temperature drop across the interlayer (see Supplementary Fig. 7). In Fig 4d, we plot cumulative thermal resistance between the GaN layer and underlying non-native substrates against layer thickness given in the literature, as a benchmarking exercise. Clearly, compared to most GaN-on-Si grown with conventional buffers, our structure possesses almost an order-of-magnitude smaller thermal resistance between the GaN layer and the silicon. In fact, the measured $TBR_{eff}$ is among the lowest values for heteroepitaxial GaN across all substrates.

To date, efforts to improve the heat extraction of GaN-on-Si HEMTs have focused primarily on two post-epitaxy pathways. One approach aims to bring the heat-extraction boundary closer to the NL, either by embedding microfluidics[54] in silicon with active cooling or micromachining the silicon substrate with metal-filled trench vias[55]. However, the buffer and its large thermal resistance still remain in both cases. Alternatively, all layers beneath the GaN are removed by etching, and then, high thermal conductivity diamond is deposited on the backside as a heat spreader[50,56]. However, along with unnecessary growth and then removal of nitride epilayers, the removal of buffer layers may cause problems with stress management. If required, buffer-less GaN-on-Si would be amenable to all these post-growth approaches to reduce its channel-to-heat-sink thermal resistance further, but without the current limitations.

**Room-temperature electronic properties and low-temperature magneto-transport of AlGaN/GaN heterostructures:**

Next, to create heterojunctions on our buffer-less GaN-on-Si templates, in a separate run, a ≈20 nm Al$_{0.22}$Ga$_{0.78}$N layer was grown after the optimised GaN layer. The 0004 ω-2θ XRD data of this structure in Fig. 5a show well-defined satellite peaks around GaN and AlGaN peaks, corroborating a sharp interface between the two top layers. XRD RSM confirmed that the Al$_{0.22}$Ga$_{0.78}$N layer is pseudomorphically strained to the underlying GaN (see Supplementary Fig. 8).

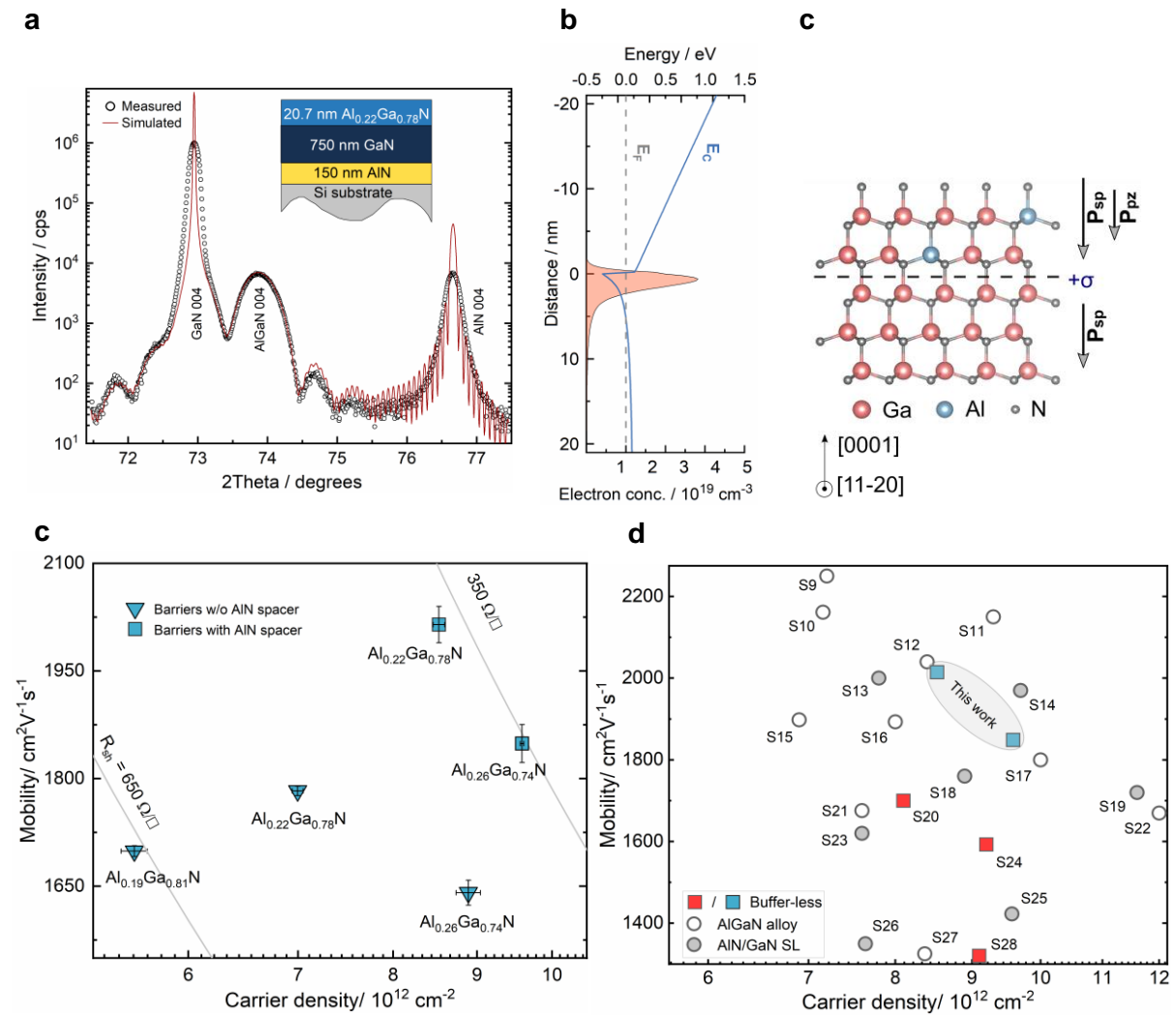

Figure 5. Interfacial confinement and room-temperature (RT) transport properties of the 2DEG. (a) Close agreement between measured XRD ω-2θ data of the symmetric (004) reflection with simulation confirms the growth and abruptness of the heterojunction (the simulated structure is shown in the inset). (b) Simulated conduction band edge of the as-grown type-I structure shows that the distribution of electrons is primarily confined on the GaN side of the interface. A schematic of the Ga-polar lattice is shown with the direction of the polarisation fields. (c) Hall-effect measured variation in 2DEG electron density and mobility with barrier properties for designs without and with the AlN spacer. The error bars represent standard error the mean. (d) Benchmark plot showing the comparison with previously reported Hall-effect values for GaN-on-Si epistructures with buffer-less and various buffer schemes (references for the data are listed in Supplementary Table 2) confirms that the 2DEGs studied in this work are among the best with simultaneous high mobility and carrier density.

This biaxial strain state in $Al_{0.22}Ga_{0.78}N$ should generate a piezoelectric polarization ($\boldsymbol{P_{pz}}$) which is further supplemented by the difference in spontaneous polarization ($\boldsymbol{P_{sp}}$) between $Al_{0.22}Ga_{0.78}N$ and GaN. As shown in the simulated energy band diagram (Fig. 5b), this polarization sheet charge at the interface[4] along with the heterojunction's large conduction band offset should confine electrons as a 2DEG in GaN.

This was verified by low-field Hall-effect measurements of this heterostructure, at room temperature (RT) which showed an average charge carrier density ($n_s$) of $7.0\times10^{12}$ /cm$^2$ and mobility ($\mu$) of 1783 cm$^2$/(V·s) along with a negative Hall coefficient. Moreover, an almost constant carrier density along with a continuously increasing mobility with the reduction in temperature (see Supplementary Fig. 9), confirmed the well-known signatures of a 2DEG.

For 2DEGs, mobility remains the key evaluator of electrical cleanliness[57] and requires careful optimisation. To test the variation of $\mu$ with $n_s$ for our structures, we grew additional samples with barrier compositions of $Al_{0.19}Ga_{0.81}N$ and $Al_{0.26}Ga_{0.74}N$. Among these three samples (triangular symbols in Fig. 5c), $n_s$ showed a progressive increase from $5.6\times10^{12}$ /cm$^2$ to $8.9\times10^{12}$ /cm$^2$ with increasing AlN mole fraction (x), consistent with increased spontaneous and piezoelectric polarization contributions[4]. The fact that we observe a lower electron mobility for x = 0.26 compared to x = 0.22 suggests that scatterers which are more screened[59] at higher carrier densities such as charged dislocations or background ionized impurities, are not dominant for x $\geq$ 0.22 barriers on our template. Instead, in this carrier density regime, increased interface roughness and alloy scattering is likely the cause for mobility reduction as the wavefunction penetrates deeper into the barrier[60]. To alleviate this penetration, we introduced a $\approx$1 nm AlN spacer between the channel and the AlGaN barrier. For samples with such a spacer and with barrier compositions of $Al_{0.22}Ga_{0.78}N$ and $Al_{0.26}Ga_{0.74}N$ we observed a significant increase in mobility to 2015 cm$^2$/V·s and 1849 cm$^2$/V·s (square symbols in Fig. 5c), respectively. Simultaneously, $n_s$ increased to $8.5\times10^{12}$ /cm$^2$ and $9.6\times10^{12}$ /cm$^2$ due to a polarization boost from the spacer. A comparison of these samples with other reported AlGaN/GaN HEMT heterostructures on silicon is shown in Fig. 5d. The overview shows that our RT Hall-effect mobilities are significantly higher than those previously achieved without buffers. Also, along with larger or comparable electron densities, our mobility values are among the highest reported for GaN-on-Si, including various buffer schemes on silicon that amass up to 8 µm of total epi-thickness (see Supplementary Table 2). These results provide clear evidence that neither buffers nor thick epilayers are needed to realise high-quality nitride heterostructures with high 2DEG mobility and density.

Low-temperature quantum magneto-transport is a more challenging metric of high-quality for 2DEGs in modulation doped[57] or undoped polar[61] heterostructures. E.g., if scattering is sufficiently low (i.e. $\omega_c\tau > 1$, where $\omega_c$ is the cyclotron frequency and $\tau$ is the carrier relaxation time), the 2DEG's longitudinal resistance ($R_{xx}$) becomes oscillatory in increasing magnetic field (i.e Shubnikov-de-Haas (SdH) oscillations are observed). The perception that delicate magneto transport requires high-quality epilayers with low defect density has compelled the nitride community to restrict such measurements primarily on samples grown on bulk GaN[62],SiC[63], or sapphire[64].The high mobilities in Fig 5 indicated that such features could be potentially observable in our structures. Hence, we investigated magneto-transport in the $Al_{0.22}Ga_{0.78}N/AlN/GaN$ structure with the highest mobility. As shown in the low-temperature normalized magnetoresistance (MR) trace of Fig. 6a, distinct SdH oscillations with progressively increasing amplitudes with the magnetic field ($B$) were seen. These oscillations appeared superimposed on a negative MR with parabolic dependence on *B*. Collectively, this resulted in a substantial change in magnetoresistance at larger fields, reaching ≈ 75 % by 14 T. Fig. 6b shows the data after removing the $B^2$-dependent background by double differentiation of the raw MR data[65]. In addition to the main SdH oscillations which begin at ≈ 4 T, a secondary oscillation is prominent in the spectra above 10 T. The respective periodicities of the two oscillations above 10 T with $B^{-1}$ yields slightly distinct sheet carrier densities behind their origins (see Supplementary Fig. 10). The calculated total *n* was ≈8.8×$10^{12}$ /cm², which is consistent with the low-field Hall effect data for this wafer (8.5×$10^{12}$ /cm²). This agreement validates the strong 2D confinement of the channel carriers. Moreover, this proves that the spin degeneracy of the single subband in the QW was removed, yielding two spin-split subbands[63,64]($n\uparrow$ and $n\downarrow$). Its origin is likely the Rashba spin-orbit coupling effect[66] in this noncentrosymmetric system, enhanced by the built-in interfacial electric field perpendicular to the QW.

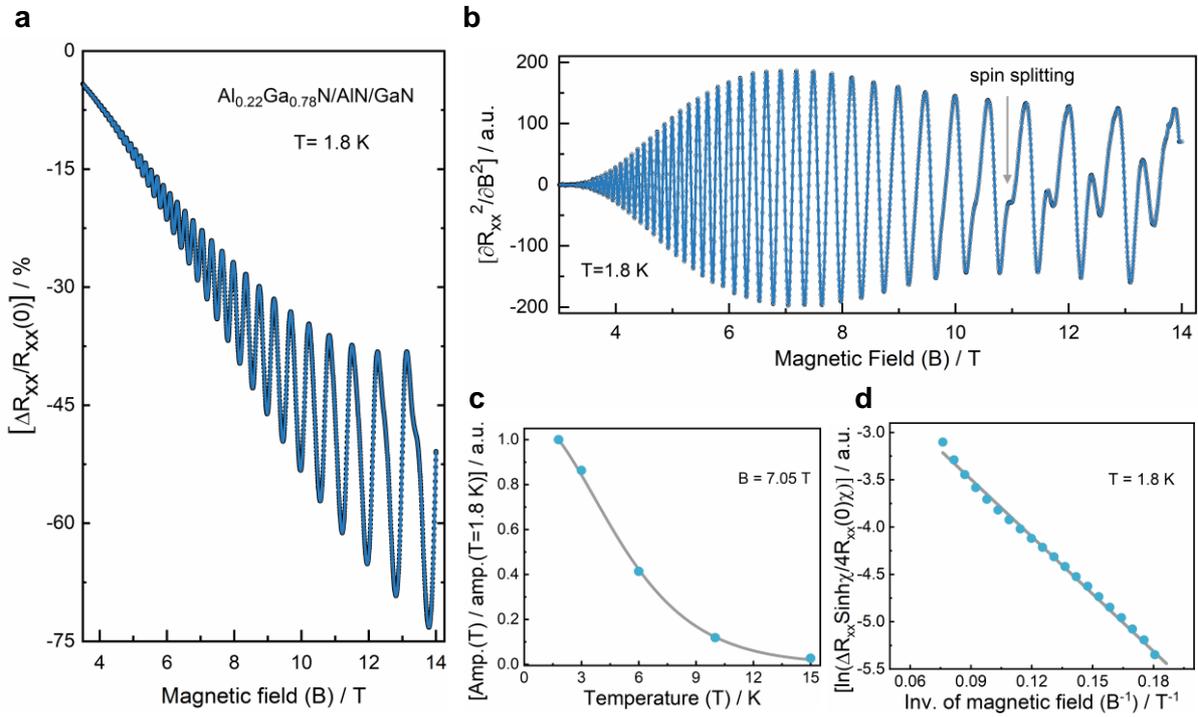

Figure 6. Quantum Hall signatures in the AlGaN/AlN/GaN 2DEG at low temperatures. (a) Change in longitudinal magnetoresistance (ΔRxx) with increasing magnetic field, B, at 1.8 K. The data has been normalised with respect to the value at *B*=0 i.e. $R_{xx}(0)$. Distinct Shubnikov de Haas (SdH) oscillations can be observed along with a parabolic background. (b) The second derivative of the raw magnetoresistance data plotted in (a) removes the background and shows only the changing SdH amplitudes with magnetic fields. After an initial increase, the peak-to-peak amplitudes starts to decrease from B ~7.5 T. Moreover, from ~10 T, a secondary oscillation is seen whose amplitude increases with B. (c) Temperature dependence of the normalised magnetoresistance peak (with respect to the value at 1.8 K) at constant magnetic field were used to extract the effective carrier mass. A representative fit (solid line) for the data (solid circles) at 7.05 T is shown. (d) Representative Dingle plot for the oscillation amplitude peaks from ~ 5.5 T to ~ 13 T and considering *m** = 0.210 $m_0$ (obtained from 6c). The straight line is a single-parameter fit to the data (solid circles) to extract the associated quantum scattering time from the slope.

For these measurements, increasing the temperatures from 1.8 K to 15 K showed a gradual damping of the SdH oscillations. This temperature dependence of the peak amplitudes was used[67] to extract the electron's effective mass (*m**) in the subband. For these fits, the peaks between 6 T to 8 T were selected as the oscillations in Fig. 6(b) were the strongest in this regime. A representative fit (for 7.05 T) is shown in Fig. 6(c). The estimated *m** values were between $(0.208 \pm 0.008)m_0$ and $(0.215 \pm 0.005)m_0$, consistent with those obtained from different techniques in the literature[68]. Also, based on these effective masses, the quantum scattering time ($\tau_q$) was extracted from the slope of the Dingle plots (Fig. 6d). $\tau_q$ is a measure of Landau level broadening and is affected by scattering events over all angles (unlike the large-angle scattering dominated transport lifetime that governs the Hall-effect mobility. Values between 0.184 ps and 0.191 ps were obtained for $\tau_q$, a relatively large lifetime compared to previous reports[69]. Effectively, this yields a quantum mobility, $\mu_q$ ($= e\,\tau_q/m^*$), of $\approx 1550$ cm$^2$/V·s.

These observations have important implications for III-nitrides. Compared to the widely investigated homoepitaxial III-arsenides and phosphides, QWs in GaN heterojunctions present a g*-factor close to that of free electron, an $m^*$ three times heavier than GaAs but comparable Rashba spin-orbit coupling. Buffer-less GaN-on-Si now presents an industrially scalable, technologically relevant platform for probing and exploiting a unique regime of mesoscopic physics.

**Conclusion and outlook:**

In summary, we quantitatively established that by controlling the coalescence of the GaN layer, the lattice and thermal mismatch challenges of III-nitride heteroepitaxy on Si can be overcome even without buffer layers. Benefiting from the absence of thermally poor conductive buffers, our structures provide an ultra-high thermally conductive pathway for heat removal. The excellent crystal quality of these buffer-less epilayers is further evidenced by the simultaneous realization of state-of-the-art 2DEG mobility and large sheet carrier density with AlGaN/GaN heterojunctions.

It is worth mentioning that thickness variations of our buffer-less design can enhance application specific functionalities. E.g., AlN provides superior channel confinement compared to AlGaN back-barriers. Hence, as a possibility we suggest that a reduction in GaN thickness could enhance this effect further, providing the short-channel immunity needed for ultra-scaled RF transistors and ultimately paving the realisation of an AlN/GaN/AlN QW-FET[70] on Si. On the other hand, for power devices, it may be possible to thicken the GaN drift layer further to increase the breakdown voltage by screening the fields from the substrate. Altogether, a new regime of exploration and implementation of lower-cost, sustainably-produced GaN-based heterostructures for high-performance HEMTs and novel quantum states in the 2D limit is envisaged.

**Materials and methods**:

**Epitaxial growth and *in situ* characterisation**:

The epitaxial structures were grown in an industrial-grade rotating-disc ($\approx 1000$ rpm) MOVPE reactor (Veeco Propel™) with trimethylgallium (7N pure, source: Pegasus chemicals), trimethylaluminum (7N pure, source: Dow chemicals), and ammonia (6N5 pure, source: BOC) as the precursors and hydrogen (source: BOC, purified to < 50 ppb through a purifier (make: Entegris)) as the carrier gas. 1 mm thick, epi-ready p-type Si(111) wafers (source: Shin-Etsu Handotai) were directly loaded into the reactor without any chemical treatment. After in situ removal of the native oxide by heating under hydrogen, the AlN NLs were grown in two temperature steps, both at pressure of 75 Torr. The first stage was grown at 750°C temperature, and after a fast temperature ramp with growth continuing, the second stage (≈140 nm) was grown at 1070°C. Before GaN growth, the temperature was ramped down to 1050°C and the reactor pressure was changed to the value intended for that sample, and these parameters were kept fixed during the entire GaN growth duration. The V/III molar ratio was kept low at ≈75 during AlN growth and ≈1500 during GaN growth. The temperature, reflectance, and curvature of the wafers were continuously measured in real time during growth with the integrated Veeco DRT-210™ process monitoring tool, which uses a narrow-bandwidth 650 nm laser. All the temperatures mentioned are emissivity corrected wafer temperatures.

**Structural and topographical characterisation**:

A ZEISS scanning electron microscope with a field emission gun (GeminiSEM 300) operated at 2 kV with 200 nA beam current was used to acquire the cross-sectional SE image of the structure. The crystalline quality of the epilayers and the epilayer-substrate alignments were measured in symmetric and skew-symmetric geometry in a 4-circle horizontal high-resolution X-ray diffractometer (Philips PW3050/65 X'Pert Pro) with monochromated CuKα$_1$ radiation and a proportional counter. The thickness of the thin AlGaN barrier layers was determined by comparing the measured ω-2θ curves with simulations obtained from the PANalytical Epitaxy software. The RSM of asymmetric XRD reflections of the epilayers were collected in a 4-circle vertical high-resolution diffractometer (Panalytical Empyrean) with monochromated CuKα$_1$ radiation and a position sensitive detector (PIXcel$^{3D}$). The sample topographies were assessed in a Bruker Dimension Icon Pro AFM in Peakforce Tapping mode using commercial SCANASYST-AIR tips (nominally having 2 nm tip radius and 0.4 N/m spring constant). The acquired raw data was processed with the Bruker NanoScope Analysis v1.9 software.

**Thermal characterisation**:

All the TTR measurements were taken at room temperature. A 10 ns, 355 nm frequency tripled Nd:YAG pump laser with a 30 kHz repetition rate was defocused on the sample surface (≈ 40 µm FWHM Gaussian profile) to induce heating and a focused CW 320 nm laser was used to monitor the reflectance. Using above-bandgap UV lasers (absorption depth < 100 nm for GaN) ensured that the heating and probing were restricted to the GaN surface. For fitting the TTR data with the model for thermal transport, thickness ($t$), material density ($\rho$), specific heat capacity ($C$), and thermal conductivity ($k$) of each layer were treated as fixed parameters. Values[27] at 298 K were used for material-dependent parameters, listed below.

Table 1. Thermal conductivity ($k$), specific heat capacity ($C$), and material density ($\rho$) for GaN, AlN, and Si. '$t$' denotes the thickness of each layer used in the investigated samples.

| Layer | $k$ / W m$^{-1}$ K$^{-1}$ | $C$ / J Kg$^{-1}$ K$^{-1}$ | $\rho$ / Kg m$^{-3}$ | $t$ / nm |
|-------|------|------|------|------|
| GaN | 140 | 415 | 6150 | 750 |
| AlN | *Fitted* | 730 | 3260 | 150 |
| Si | 149 | 700 | 2330 | $10^6$ |

**Energy band calculations**:

The energy band diagram of the $Al_{0.2}Ga_{0.8}N$/GaN heterostructure was calculated by self-consistent solution of Schrodinger equation and Poisson equation using the software BandEng (available from https://my.ece.ucsb.edu/mgrundmann/bandeng.htm). The in-built material properties within the software were used for the simulation.

**Electrical characterisation**:

The sheet resistance, carrier concentration and mobility at room temperature and low temperatures (down to 77 K) of all samples were measured in an Ecopia HMS 5000 system with a fixed magnetic field of $\pm$ 0.55 T while sourcing 50 $\mu$A to 150 $\mu$A DC current. For these measurements, square 1 $cm^2$ samples were processed by conventional photolithography with annealed Ti/Al/Ti/Au ohmic contacts at the corners to conform to square van der Pauw pattern. The average values from three tested samples of each wafer are reported in the main text. The cryogenic (down to 1.8 K) magneto-transport measurements were conducted in a Dynacool Physical Property Measurement System (PPMS) by Quantum Design Ltd. with a superconducting magnet of variable field strength (up to 14 T). Magnetoresistances were measured in a four-probe configuration with 10 $\mu$A AC current excitation using standard lock-in techniques. For all electrical characterisations under magnetic fields, the applied field was perpendicular to the heterostructure.

**Acknowledgement**

This growth and material characterisations conducted in this research was supported by the Engineering and Physical Sciences Research Council (EPSRC) under the grants 'Hetero-print': A holistic approach to transfer-printing for heterogeneous integration in manufacturing (EP/R03480X/1) and 'InGaNET': Integration of RF Circuits with High Speed GaN Switching on Silicon Substrates (EP/N017927/1). The AFM scans and the magneto-transport measurements were supported by EPSRC grants, namely, 'Cambridge Royce facilities grant' (EP/P024947/1) and 'Sir Henry Royce Institute - recurrent grant' (EP/R00661X/1). Alexander M Hinz acknowledges the Deutsche Forschungsgemeinschaft for his Research Fellowship at the University of Cambridge. D J Wallis would like to thank the support of EPSRC through grant no. EP/N01202X/2. The authors are grateful to Dr Cheng Liu for assistance with the magneto-transport measurements and to Kalyan Kasarla and Terry Devlin from Veeco for their support with the MOCVD reactor. For the purpose of Open Access, the authors have applied a CC BY public copyright license to any Author Accepted Manuscript (AAM) version arising from this submission.

**Author contributions**

S.G. and R.A.O. conceptualised the investigation. S.G. and A.M.H. did the epitaxial growths and relevant data analysis. S.G., M.F., D.J.W., and R.A.O. conducted the structural and topographical characterisations of the samples and analysed the collected data. J.W.P., D.F., and M.K. performed the thermal characterisation. S.G. carried out the Hall-effect and low-temperature magneto-transport measurements and analysed the collected data. S.G., M.F., and R.A.O co-wrote the manuscript and all the authors commented on it.

# Supplementary Material

# Buffer-less Gallium Nitride High Electron Mobility Heterostructures on Silicon

Saptarsi Ghosh[1,2,*], Martin Frentrup[1], Alexander M. Hinz[1], James W. Pomeroy[3], Daniel Field[3], David J. Wallis[1,4], Martin Kuball[3], and Rachel A. Oliver[1]

[1]Department of Materials Science and Metallurgy, University of Cambridge, Cambridge CB3 0FS, United Kingdom

[2]Department of Electronic and Electrical Engineering, Swansea University, Swansea SA1 8EN, United Kingdom

[3]H.H. Wills Physics Laboratory, University of Bristol, Bristol BS8 1TL, United Kingdom

[4]Centre for High Frequency Engineering, Cardiff University, Cardiff CF24 3AA, United Kingdom

*E-mail: saptarsi.ghosh@swansea.ac.uk

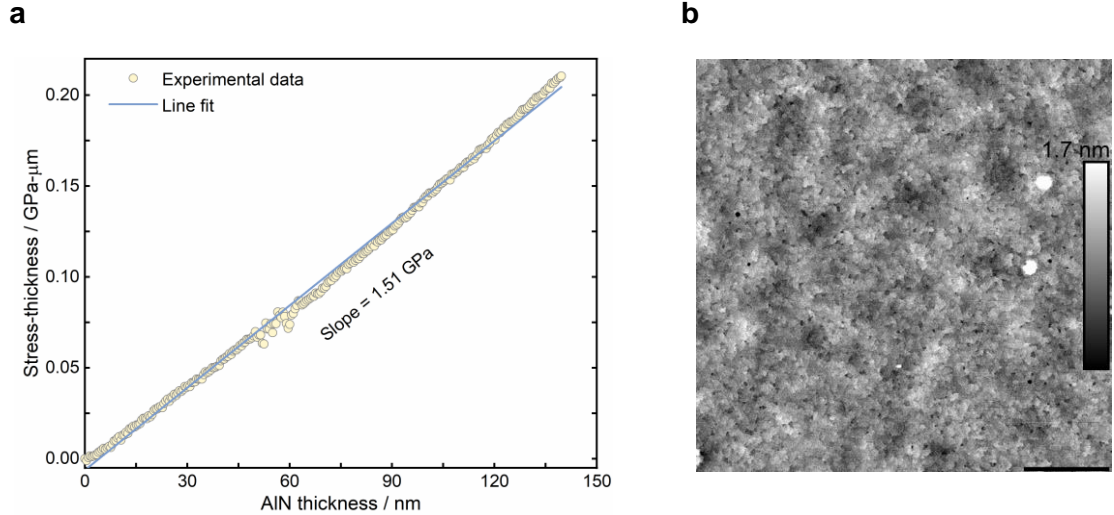

Supplementary Figure S1. (a) Representative stress-thickness curve and (b) topography of the AlN nucleation layers (NL). The scale bar in the image is 1 μm.

Figure S2 (a) shows the evolution of the stress-thickness with increasing thickness during the second stage (≈ 140 nm) of the AlN layer growth for one of the GaN/AlN/Si hetero-structures. The data points (solid symbols) could be fitted with a straight line having a positive slope, which suggests that the stress was constant and tensile. For the five GaN growths shown in Fig. 1a of the main manuscript, the mean stress during the underlying AlN NL growth was $(1.46 \pm 0.04)$ GPa.

Figure S3 (b) shows the surface topography of a representative 5 μm × 5 μm area of an identically grown AlN NL on Si. AFM scans from different regions of the wafer showed sub-nm rms roughness with a mean value of $(0.32 \pm 0.03)$ nm.

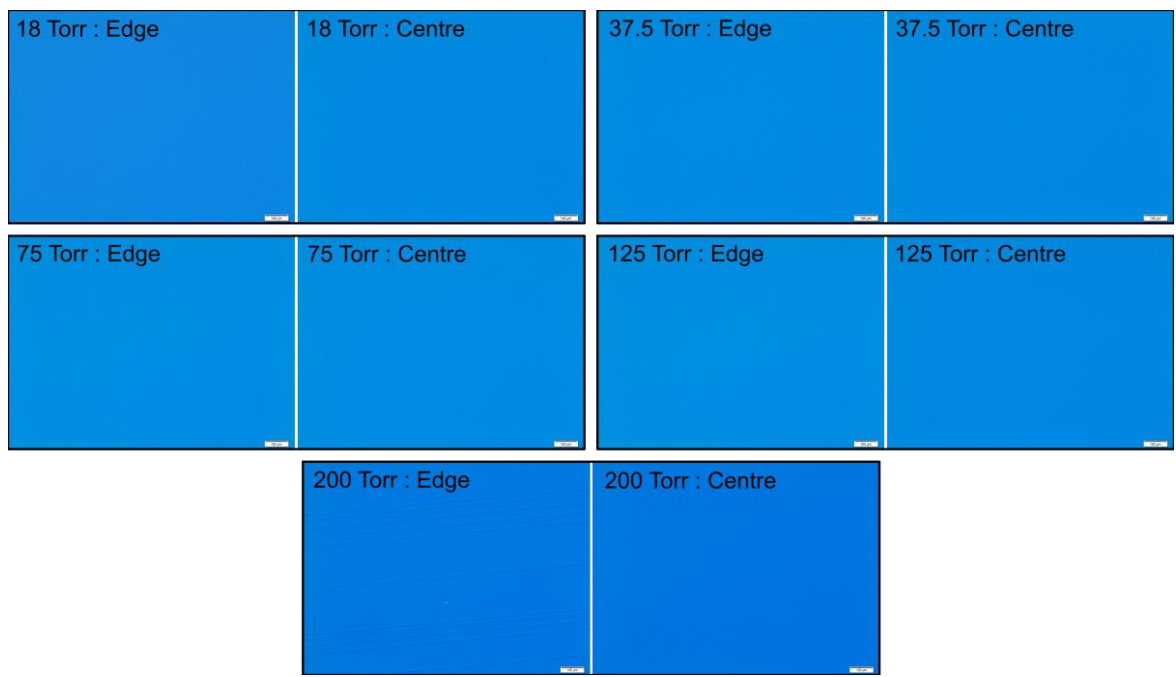

Supplementary Figure S2. Large-area Differential Interface Contrast (DIC) optical micrographs of the surfaces after ≈ 800 nm of GaN growth at different reactor pressures.

Figure S2 shows large-area Differential Interface Contrast (DIC) optical micrographs of the surfaces after ≈ 800 nm of GaN growth at different reactor pressures. Representative images were acquired near the wafers' edge (≈ 1 cm inside) and at the centre. All the surfaces show pit-free mirror-smooth morphology, indicating complete coalescence of these submicron heteroepitaxial layers. The first appearance of cracks can be observed near the edge of the wafer for which the GaN layer was grown at 200 Torr reactor pressure, though they do not propagate to the wafer centre. For growth pressures below this, all the wafers are crack-free. The scale bars in all the images are 100 $\mu$m.

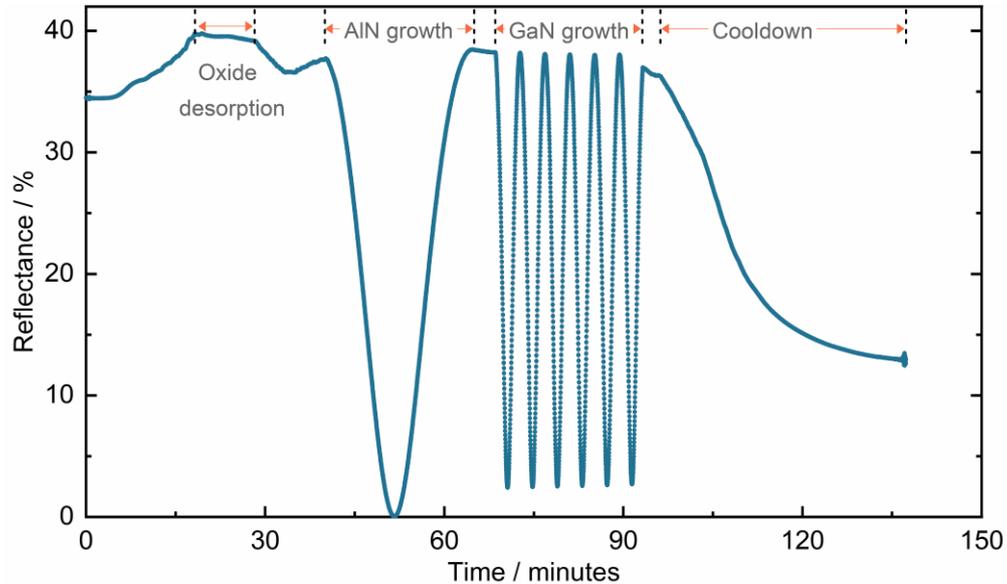

Supplementary Figure S3. Evolution of reflectance during a representative growth run of 800 nm GaN/ 150 nm AlN on a silicon(111) wafer. The growth steps of in-situ oxide desorption, AlN growth, GaN growth, and post-growth cooldown are indicated.

Figure S3 shows the evolution of reflectance during a representative growth run of 800 nm GaN/ 150 nm AlN on a silicon (111) wafer. For this sample, the GaN layer was grown at 75 Torr. The duration of *in situ* oxide desorption, AlN growth (the second stage i.e. ≈ 140 nm thick part), GaN growth, and post-growth cooldown steps are marked. The silicon wafer is loaded directly into the growth chamber without any chemical or thermal treatment beforehand. As seen, after loading (t = 0 s), this entire growth run is completed within 2.5 hours. In contrast to GaN, the growth of AlGaN alloy layers suffers from additional gas-phase pre-reactions and much slower growth rates. E.g., of the various step-graded AlGaN buffer designs described in [Ref. 1], the growth times for the fastest 1.0 μm and 1.7 μm buffer layers themselves were ≈ 2 hours and ≈ 3.3 hours, respectively. On the other hand, AlN/GaN superlattice (SL) buffers require periodic changes in growth conditions for each layer of the SL. From [Ref. 2], it can be found that growth of 1.4 μm to 1.8 μm buffers require additional growth times of ≈ 1.5 hours. Thus buffer-less structures are considerably beneficial in terms of thermal budget as well as metal-organic precursors and ammonia usage.

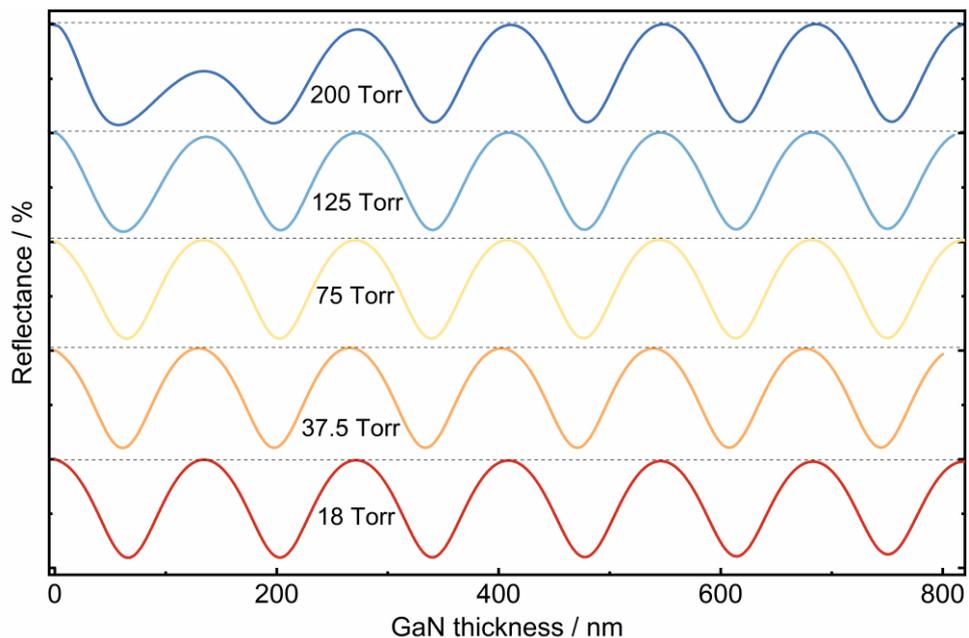

Supplementary Figure S4. Variation of real-time reflectance transients with the reactor pressure during the growth of the ≈ 800 nm thick GaN layers.

Figure S4 shows real-time reflectance transients, which were measured during the growth of the ≈ 800 nm thick GaN layers at different reactor pressures. The dashed horizontal line marks the value of peak reflectance of the fifth complete oscillation. For a growth pressures of 75 Torr and below, the reflectance achieves this value by the first complete oscillation peak. However, for higher pressures, increasingly further progression into growth is required. All the oscillations eventually have the same peak-to-peak reflectance. The reflectance at the beginning of the GaN growth was identical for all the growths, and this data has been offset for comparison. For the 650 nm laser wavelength used for these measurements, each complete oscillation corresponds to a thickness of ≈ 137 nm for GaN.

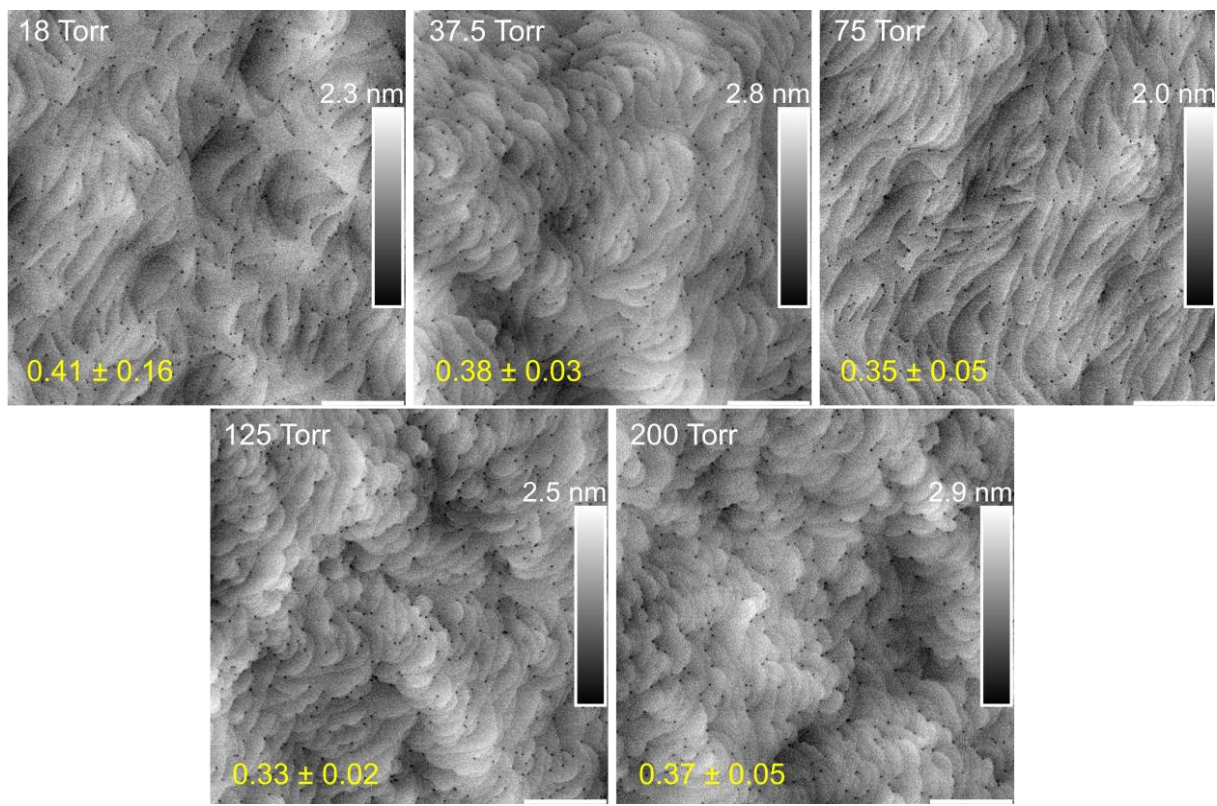

Supplementary Figure S5. Topography of the ≈ 800 nm thick GaN layers grown at different pressures. The scale bar in all the 25 µm² images is 1 µm. Root mean square (rms) roughness values in nm are given in yellow on each image.

The topography of the ≈ 800 nm thick GaN layers grown at different pressures is shown in Figure S5. All the 5 µm × 5 µm surfaces show a morphology typical of MOCVD-grown GaN with steps pinned at threading dislocations (which appear as tiny pits in the images). The sub-nm mean rms roughness for all the wafers (annotated on the images) indicates similarly smooth topography regardless of the growth pressure.

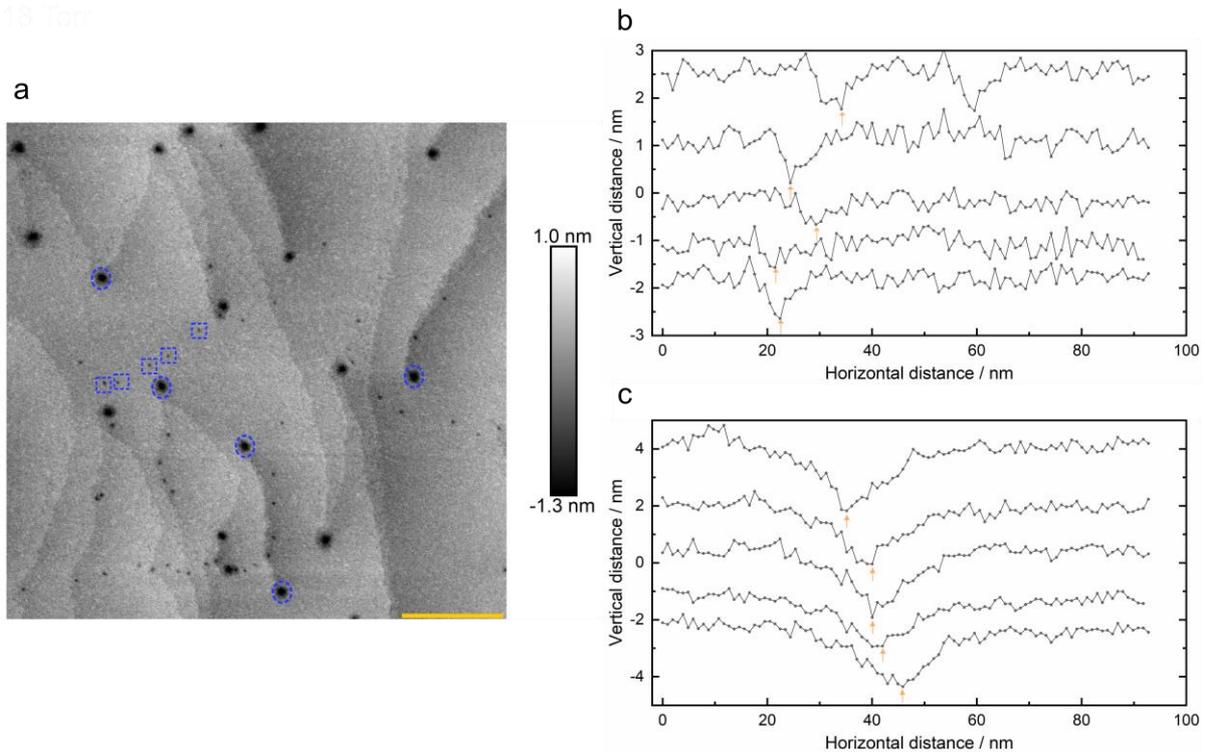

Supplementary Figure S6. Dislocation quantification from high-resolution AFM scans of ≈ 800 nm thick GaN samples. (a) A representative 1 μm × 1 μm surface topography of a GaN/AlN/Si wafer and (b)/(c) waterfall plot of line profiles extracted from the AFM dataset. The scale bar in (a) is 200 nm. The arrows in (b) and (c) denotes the maximum measured depth for each profile.

For the dislocation quantification of the ≈ 800 nm thick GaN samples, high-resolution AFM scans have been recorded. Figure S6 (a) shows such a representative 1 μm × 1 μm AFM image of the surface topography of a GaN/AlN/Si wafer in which the GaN layer was grown at 37.5 Torr (scale bar: 200 nm). For each investigated wafer, nine such random areas were scanned for statistics. All the images were acquired with high sampling density to identify the small pits created from surface terminations of threading dislocations. The observed pits created due to dislocations could be categorized as either shallower and narrower or deeper and wider. These were assigned to have either edge or screw characters, respectively. Five of such dislocations with edge character are marked with open squares in (a) and their line profiles are shown in (b). Also, the line profiles five dislocations with screw characters (marked with open circles in (a)) are shown in (c).

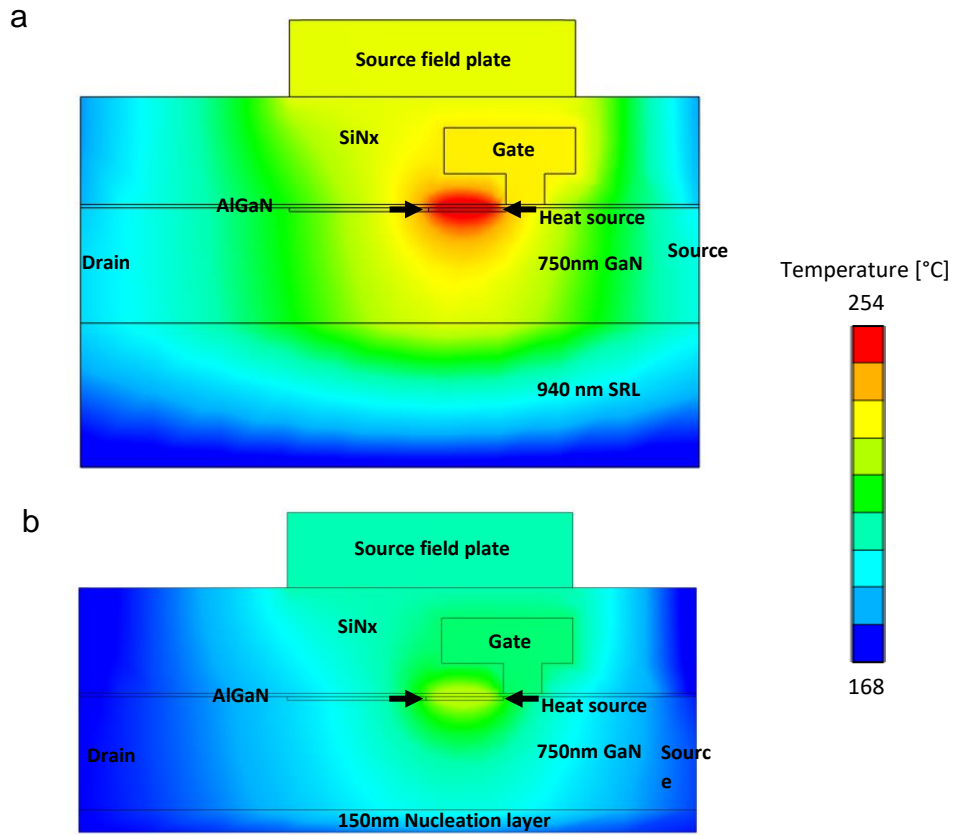

Supplementary Figure S7. 3-D FEM thermal simulations of 10×350 µm, 60 µm gate pitch AlGaN/GaN HEMTs on 675 µm-thick silicon substrates with (a) conventional buffer and (b) buffer-less design.

Figure S7 shows 3-D FEM (Ansys) thermal simulations of $10 \times 350$ µm, 60 µm gate pitch AlGaN/GaN HEMTs on 675 µm-thick silicon substrates with (a) conventional buffer and (b) buffer-less design. For the simulations, a power dissipation of 5 W/mm within an 0.5 µm-long region at the drain edge of each gate is considered. The bottom surface of the substrate (not shown) is fixed at 22°C. (a) shows the temperature gradient around the channel heat source in the middle of the central gate finger where the layers between GaN and substrate is replaced with a strain relief layer (SRL) with a $TBR_{eff}$ equivalent to 109 $m^2K$/GW [Ref. 3]. (b) shows our buffer-less design with a 150 nm-thick nucleation layer (as evaluated, $TBR_{eff} = 11$ $m^2K$/GW). As both plots use the same temperature scale for comparison, it is evident that the temperatures in the buffer-less structure throughout are substantially lower. For example, the $\Delta T$ across the SRL and nucleation layers shown in (a) and (b) are 56°C and 15°C, respectively.

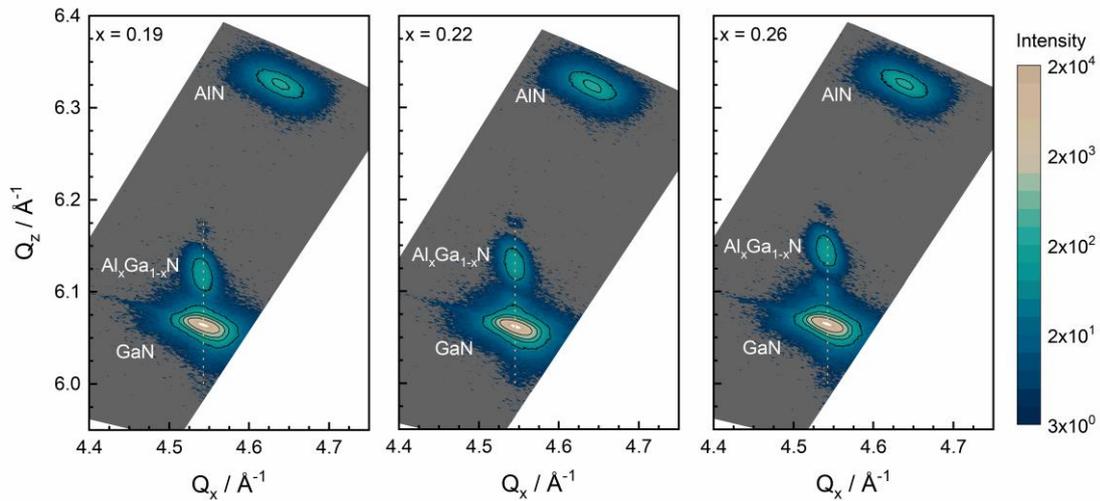

Supplementary Figure S8. XRD reciprocal space maps around the AlN 20-25 reflection for $Al_xGa_{1-x}N$/GaN/AlN heterostructure samples with different AlN mole fractions ($x$ = 0.19, 0.22, and 0.26).

Figure S8 shows XRD reciprocal space maps around the AlN 20-25 reflection for $Al_xGa_{1-x}N$/GaN/AlN heterostructure samples with different AlN mole fractions (x = 0.19, 0.22, and 0.26). For all compositions, the centroid of the reciprocal lattice spot of the barrier layer is vertically aligned with the GaN layer (indicated by dashed lines). Also, as x increases, the AlGaN reflection progressively moves to larger $Q_z$ values. These features confirm that for all the studied $Al_xGa_{1-x}N$ compositions, the barriers are pseudomorphically strained to the GaN layers underneath.

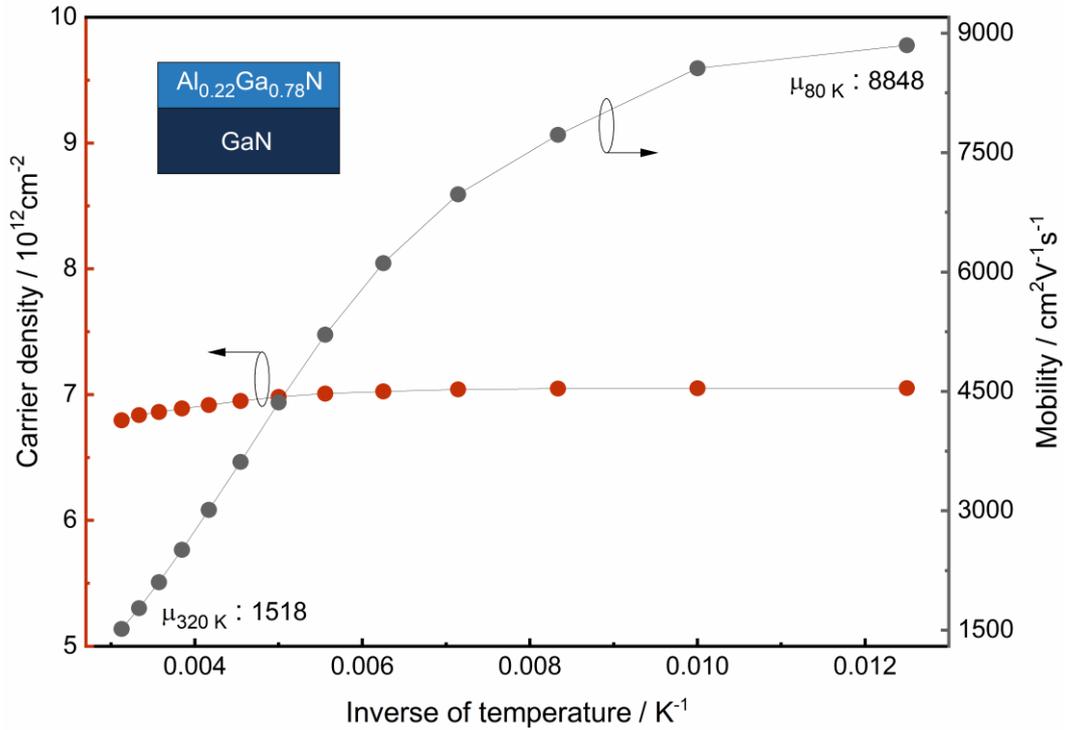

Supplementary Figure S9. Temperature-dependent carrier density and Hall-mobility for a sample with $Al_{0.22}Ga_{0.78}N$ barrier.

Figure S9 shows the carrier density and Hall-mobility for a sample with $Al_{0.22}Ga_{0.78}N$ barrier as function of inverse temperature. Between $0.003 K^{-1}$ (320 K, $n_s$: $6.8 \times 10^{12} cm^{-2}$) and 0.0125 $K^{-1}$ (80 K, $n_s$: $7.1 \times 10^{12} cm^{-2}$), the carrier density stays nearly constant. In contrast, the mobility continuously rises with a reduction in temperature and reaches 8848 $cm^2 V^{-1} s^{-1}$ at 80 K for this sample. This effectively caused a $\approx 6\times$ reduction in the sheet resistance from 605 to 100 $\Omega/\square$ (not shown). The trendlines are only for a guide to the eye.

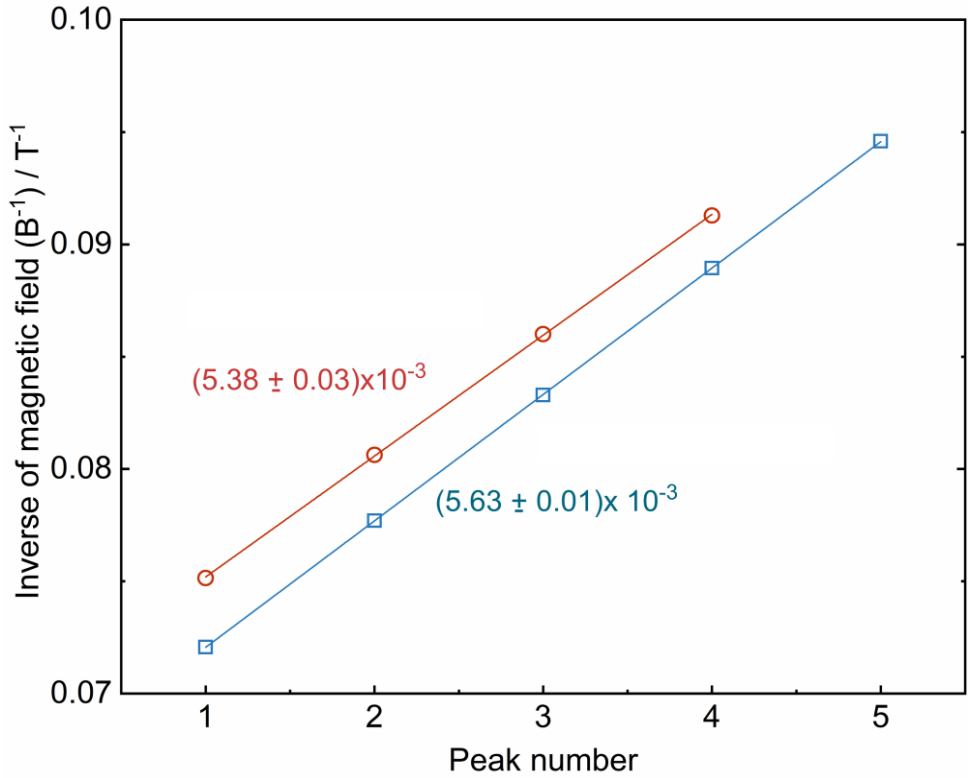

Supplementary Figure S10. Position of the successive peaks for the two distinct oscillations observed in the magnetotransport data above 10 T at 1.8 K.

The low-temperature (1.8 K) magneto-transport data presented in the main manuscript show single peaks for the oscillations below 10 T and double peaks for the oscillations above 10T (compare Fig. 6 (b) in the main manuscript). Figure S10 above shows the position of these two successive peaks for the distinct oscillations observed above 10 T. The straight lines are linear fits to the experimental data (open symbols). The extracted slopes ($\Delta B^{-1}$) along with their standard error are annotated for each oscillation. For SdH oscillations of carriers in a single sub-band, the carrier density is given by ($2e \times \Delta B^{-1}/h$). As these oscillations arise due to the spin splitting of the same sub-band, the formula for the associated carriers should be ($e \times \Delta B^{-1}/h$) i.e. without the spin degeneracy factor of 2. This yields spin-up ($n\uparrow$) and spin-down ($n\downarrow$) densities $\approx 4.47 \times 10^{12}$ cm$^{-2}$ and $\approx 4.31 \times 10^{12}$ cm$^{-2}$, respectively.

Supplementary Table 1. Summary of effective thermal boundary resistances (TBR$_{eff}$) between heteroepitaxial GaN and the various non-native substrates plotted in Fig. 4d of the main manuscript. Sample ID refers to the symbol used for that article in the main text and reference denotes the reference number in this supplementary material. The interlayer type and its thicknesses are also listed if this information is available.

| Substrate | Sample-ID [Reference] | Interlayer type | Thick-ness / nm | TBR$_{eff}$ / m$^2$ K GW$^{-1}$ | Note |
|---|---|---|---|---|---|
| Sapphire | S1 [4] | AlN | 40 | 300 | The mean value was plotted from upper and lower bounds. |
| Sapphire | S2 [5] | - | 30 | 120 | The exact type was interlayer was not mentioned. |
| SiC | S2 [5] | - | 30 | 33 | The exact type was interlayer was not mentioned. |
| SiC | S3 [3] | AlN | 30 | 14 | Value was extracted from a figure in the article. |
| SiC | S4 [6] | AlN | 40 | 120 | Thickness was not explicitly mentioned. We assumed it to be 40 nm. |
| SiC | S5 [7] | AlN | 90 | 25 | - |
| SiC | S5 [7] | 24% AlGaN | 3 | 20 | Not shown in our plot. |
| SiC | S6 [8] | AlN | 40 | 36.8 | Value was extracted from a figure in the article. In the article, TBR$_{eff}$ was shown to increase with temperature. We plotted the value for the lowest temperature. |
| SiC | S6 [8] | AlN | 80 | 25.8 | Same as above. |
| SiC | S6 [8] | AlN | 30 | 35 | Same as above. |
| SiC | S7 [9] | AlN | 40 | 8.4 | Same as above. |
| SiC | S7 [9] | AlN | 70 | 20 | Same as above. |
| SiC | S7 [9] | AlN | 70 | 24.2 | Same as above. |
| Si | S2 [5] | AlN and strain-relief layer | 1000 | 33 | The exact type of the strain-relief layer was not mentioned. As its thickness was also not given, we assumed it to be 1000 nm. |
| Si | S3 [3] | Several layers of AlGaN | 940 | 109.3 | This plotted value is only for the buffer. Including the NL and the associated interfaces, total thermal resistance would be higher. The compositions of the AlGaN were not specified. |
| Si | S4 [6] | LT AlN +GaN+ LT AlN | 487 | 70 | - |
| Si | S8 [10] | AlN/GaN SL only | 780 | 112 | - |
| Si | S8 [10] | AlN | 100 | 5.3 | - |
| Si | S8 [10] | AlN | 100 | 7 | - |
| Si | This work | AlN | 150 | 11 | - |

Supplementary Table 2. Electron density (ns) and mobility ($\mu$) of the various AlGaN/GaN-based high electron mobility 2DEGs plotted in Fig. 5d of the main text. For consistency, only Hall-effect measured values at room-temperature are considered for comparison. The specification of the MOCVD-grown epilayers on Si(111) for each sample is also noted.

| Sample-ID [Reference] | Wafer dia. / inch | $t_{AlN}$ / nm | Buffer type [Thickness / nm] | $t_{GaN}$ / nm | Total epi.* / nm | $n_s$ / $10^{12}$ cm$^{-2}$ | $\mu$ / cm$^2 \cdot V^{-1} \cdot s^{-1}$ |
|---|---|---|---|---|---|---|---|
| S9 [11] | 6 | 150 | Graded AlGaN [1700] | 2750 | 4600 | 7.2 | 2250 |
| S10 [12] | 4 | 284 | Graded AlGaN [1155] | 1250 | 2689 | 7.2 | 2161 |
| S11 [13] | 4 | 270 | AlGaN single layer [330] | 2000 | 2600 | 9.3 | 2150 |
| S12 [14] | 4 | 270 | AlGaN single layer [330] | 3000 | 3600 | 8.4 | 2040 |
| S13 [15] | 8 | NA | AlGaN/AlN multilayer [NA] | NA | 5000 | 7.8** | 2000 |
| S14 [16] | 6 | NA | NA [NA] | NA | 4500 | 9.7 | 1970 |
| S15 [17] | 6 | 100 | Graded AlGaN [600] | 500 | 1200 | 6.9 | 1898 |
| S16 [17] | 6 | 100 | Graded AlGaN [1000] | 500 | 1600 | 8.0 | 1893 |
| S17 [18] | 6 | 193 | Graded AlGaN [976] | 1260 | 2429 | 10.0 | 1800 |
| S18 [19] | NA | NA | NA [2000] | 6000 | 8000 | 8.9 | 1760 |
| S19 [20] | 8 | 150 | AlGaN single layer with superlattice [2050] | 1000 | 3200 | 11.6 | 1720 |
| S20 [21] | 2 | 200 | Buffer-less | 300 | 500 | 8.1# | 1700 |
| S21 [17] | 6 | 100 | Graded AlGaN [1000] | 500 | 1600 | 7.6 | 1676 |
| S22 [22] | 4 | 180 | AlGaN single layer [2300] | 150 | 2630 | 12.0## | 1670## |
| S23 [23] | NA | NA | AlGaN based buffer [4500] | 200 | 4700 | 7.6 | 1620 |
| S24 [24] | 8 | 130 | Buffer-less | 950 | 1080 | 9.2 | 1593 |
| S25 [25] | 4 | 100 | AlGaN single layer with Superlattice [ $\geq$ 4000] | $\geq$ 500 | 5500 | 9.6** | 1423 |
| S26 [26] | 4 | 80 | AlGaN single layer with Superlattice [3030] | 1000 | 4110 | 7.6 | 1350 |
| S27 [27] | 2 | 100 | Graded AlGaN [1000] | 500 | 1600 | 8.4 | 1325 |
| S28 [28] | 2 | 200 | Buffer-less | 500 | 700 | 9.1 | 1320 |
| This work | 6 | 150 | Buffer-less | 800 | 950 | 8.5 | 2014 |
| This work | 6 | 150 | Buffer-less | 800 | 950 | 9.6 | 1849 |

*Total epilayer thickness excludes the barrier thickness (usually $\approx$ 20 nm)

# Same sample before passivation showed $n_s$ to be $4.9 \times 10^{12}$ cm$^{-2}$
*$n_s$ not given, calculated from $R_{sheet}$ and $\mu$
##after substrate removal